# Evolution of small defect clusters in ion-irradiated 3C-SiC: combined cluster dynamics modeling and experimental study

C. Liu[a], L. He[b], Y. Zhai[b], B. Tyburska-Püschel[a], P. M. Voyles[b], K. Sridharan[a], D. Morgan[a,b], I. Szlufarska[a,b]

[a] University of Wisconsin-Madison, Department of Engineering Physics, 1500 Engineering Dr., Madison, WI, 53706, U.S.A.

[b] University of Wisconsin-Madison, Department of Material Science and Engineering, 1509 University Ave., Madison, WI, 53706, U.S.A.

**Abstract**

Distribution of black spot defects and small clusters in 1 MeV krypton irradiated 3C-SiC has been investigated using advanced scanning transmission electron microscopy (STEM) and TEM. We find that two thirds of clusters smaller than 1 nm identified in STEM are invisible in TEM images. For clusters that are larger than 1 nm, STEM and TEM results match very well. A cluster dynamics model has been developed for SiC to reveal processes that contribute to evolution of defect clusters and validated against the (S)TEM results. Simulations showed that a model based on established properties of point defects (PDs) generation, reaction, clustering, and cluster dissociation, is unable to predict black spot defects distribution consistent with STEM observations. This failure suggests that additional phenomena not included in a simple point-defect picture may contribute to radiation-induced evolution of defect clusters in SiC and using our model we have determined the effects of a number of these additional phenomena on cluster evolution. Using these additional phenomena it is possible to fit parameters within physically justifiable ranges that yield agreement between cluster distributions predicted by simulations and those measured experimentally.

# 1 Introduction

SiC is an attractive material for nuclear energy applications because of its advantageous properties such as high-temperature strength, low neutron cross section, excellent stability under corrosion and oxidation, good thermal conductivity and low thermal expansion coefficient [1, 2]. In many of the reactor applications, SiC or SiC-SiC composites are subject to irradiation, which can lead to deterioration of some or all of the above properties. In particular, radiation can induce undesirable swelling, which in turn can lead to creep and microcracking in this material. Swelling occurs due to accumulation of radiation-induced defects, including PDs, small interstitial clusters, large dislocation loops, and vacancy clusters. The exact nature of defects that contribute to swelling depends strongly on the temperature. For example, black spot defects (BSDs) and small dislocation loops are the dominant defects at relatively low temperatures (≤1073 K), vacancies do not become mobile until temperatures as high as 1273-1473 K where they can form voids, and Frank faulted loops and dislocation networks are generally not observed temperatures up to 1673 K[1, 3, 4].

Here, we are interested in a temperature range relevant to light water reactors (548 – 598 K [5]). Many studies have been dedicated to understanding distributions of defects in this regime and their contributions to swelling and creep [6, 7]. For instance, Katoh *et al.* [6] used transmission electron microscopy (TEM) to measure the distribution of defect clusters in irradiated SiC and used this information to estimate the expected swelling. Interestingly, only 10–45% of the macroscopically measured swelling was accounted for by the clusters and loops that were visible in TEM. One of the outstanding challenges in this field is to be able to measure and quantify a distribution of very small defect clusters that are difficult to measure in traditional TEM. These defects are often referred to as BSDs because of their appearance in TEM bright field images, if they are visible at all. These defects are often assumed to be small interstitial clusters because in this temperature range vacancies are immobile [3] (migration barriers for Si and C vacancies are 2.4 and 3.67 eV, respectively[8]) and therefore vacancies can hardly form extended defect clusters.

In this study, high resolution scanning transmission electron microscopy (HR STEM) is used to identify the presence of BSDs present in cubic (3C) SiC irradiated with 1 MeV Kr$^+$. We also report on developments of a cluster dynamics (CD) model that describes evolution of irradiation-induced defects. This model is then used to predict cluster size distribution in irradiated SiC, which is then compared to those measured experimentally. We find that a model based on all the known defect properties in SiC is not capable of producing a reasonable agreement between simulations and experiments. We propose additional physical phenomena that are likely present in irradiated SiC and discuss their effects on the cluster size distribution with arguments that close the gap between simulation and experimental results.

# 2 Methods
## 2.1 Experimental procedure

3C-SiC for this research produced from Rohm and Haas Company was polycrystalline with a typical grain size of 5 μm. It was chemo-mechanically polished down to the surface roughness of about 20 nm (mirror finish). The samples were then irradiated by 1 MeV Kr ions at the Frederick Seitz Materials Research Laboratory Center of University of Illinois Urbana-Champaign (UIUC) at 800 °C up to fluences of $3\times10^{14}$ Kr/cm$^2$ at the flux of $3\times10^{12}$ Kr/(cm$^2$s). The estimated peak damage level was 0.4 dpa at 0.3 μm below surface, assuming the displacement energies of 20 eV for C and 35 eV for Si (see **Figure 1**). More details of these calculations are given in Ref. [9].

TEM cross sectional samples were prepared by mechanical wedge polishing with diamond lapping films, followed by ion milling using Fischione 1050 at 3-4 keV, with final milling of 0.6 keV for 10 minutes to reduce ion-milling damage. TEM imaging was conducted on a FEI Tecnai TF30 operated at 300 keV, and STEM observation was performed at 200 keV in a probe $C_s$–corrected FEI Titan S-Twin. The incident electron beam was parallel to the <110> axis of selected grains. TEM phase-contrast images were acquired with collection angles 0 to 40 mrad. Low-angle annular dark-field (LAADF) STEM images were acquired with a 17.5 mrad probe semi-convergence angle and 18.2 mrad – 91 mrad collection angle. Imaging regions in the cross sectional samples were around 0.1- 0.2 μm deep from the surface with a damage level of 0.24 - 0.32 dpa. The thickness of imaging areas was determined using low-loss electron energy loss spectra (EELS) and an inelastic mean free path of 166 nm for 300 keV electrons in SiC, estimated using the methods in Ref. [10, 11]. Particle density was calculated as the number of particles per unit specimen volume, i.e., the product of imaging areas and sample thickness.

## 2.2 Cluster dynamics simulations

Cluster dynamics (CD) has been widely used to describe such processes as interstitial clustering under irradiation [12, 13], solution component precipitation during thermal aging [14], and phase transformation during annealing [15]. Common assumptions and characteristics of CD models are: (i) clusters are distributed in a gas of their components, such as single interstitials, solvent, and primary phase components, and (ii) clusters are assumed to be in a mean field environment and therefore there is no spatial information considered. The main prediction of CD models is the size distribution of clusters and its evolution as a function of time. Time evolution of the concentration of clusters of size *n* is described by the following master equations

$$\frac{dC_n}{dt} = P_n + J_{n-m \to n} - J_{n \to n+m} + Q_n \qquad \text{Equation 1}$$

In this equation, $P_n$ is the production rate of clusters during irradiation, $J_{n-m \to n}$ and $J_{n \to n+m}$ are the net fluxes that lead to an increase in the cluster size, $Q_n$ is the reaction rate of cluster *n* with other kinds of reactants (e.g., surface, grain boundaries etc.), except other clusters. A detailed discussion of how we determine the above parameters is provided later in this paper. The CD master equations were solved numerically using the Suite of Nonlinear and Differential/Algebraic equations Solvers (SUNDIALS) developed by Lawrence Livermore National Laboratory (LLNL)[16].

In our model six kinds of PDs that can be directly introduced by a cascade [17] are considered. These are carbon interstitial ($C_I$), silicon interstitial ($Si_I$), carbon vacancy ($V_C$), silicon vacancy ($V_{Si}$), carbon antisite ($C_{Si}$), where a carbon interstitial takes the original place of a silicon atom, and silicon antisite ($Si_C$). In addition, a relatively stable complex $V_C - C_{Si}$ can be formed due to instability of $V_{Si}$, as first proposed by Bockstedte et al. [18]. The migration energy barriers for interstitials and vacancies have been taken from previously reported density functional theory (DFT) calculations [19]. These are 0.67 eV for $C_I$, 0.89 eV for $Si_I$, 3.66 eV for $V_C$ and 2.4 eV for $V_{Si}$ [8]. Diffusion coefficients for antisite defects were calculated using the method described in Ref. [8]. There were no voids observed in our experimental samples, which is consistent with previous studies that showed that void swelling starts in the temperature range of 1100 °C– 1250 °C [4]. This result is consistent with the fact that vacancies are expected to be immobile at the temperatures of our experiments. Given this lack of vacancy mobility, it is reasonable to assume that in our experiments clusters can only form by absorbing and emitting interstitials. Similarly, as in the study by Huang et al. [20], we assume all clusters to be stoichiometric. In addition, clusters are assumed to lie on {111} planes and have circular morphologies. This is consistent with several TEM experiments and analysis conducted by Snead et al.[21], Katoh et al. [4] and Yano et al. [22]. We assume here that defect reactions follow the so-called recombination model proposed by Swaminathan et al. [8], which means that we are accounting for known barriers to PD reactions as calculated from DFT.

Under the above assumptions, master equations that govern evolution of defects in our system can be written as

For clusters:
$$\frac{dC_n}{dt} = J_{n-1 \to n} - J_{n \to n+1} \qquad \text{Equation 2}$$

For carbon interstitials ($C_I$):
$$\frac{dC_{C_I}}{dt} = P_{C_I} - 2 \cdot x_{C_I} \cdot J_{1 \to 2} - \sum_{n \geq 2} x_{C_I} \cdot J_{n \to n+1} + Q_{C_I} \qquad \text{Equation 3}$$

For silicon interstitials ($Si_I$):
$$\frac{dC_{Si_I}}{dt} = P_{Si_I} - 2 \cdot x_{Si_I} \cdot J_{1 \to 2} - \sum_{n \geq 2} x_{Si_I} \cdot J_{n \to n+1} + Q_{Si_I} \qquad \text{Equation 4}$$

For other point defects (OPDs)
$$\frac{dC_{OPD}}{dt} = P_{OPD} + Q_{OPD} \qquad \text{Equation 5}$$

In the above equations, $\frac{dC_n}{dt}$ is the rate of change of atomic fraction of clusters consisting of $n$ atoms, $J_{n \to n+1}$ is the net flux flow between clusters adjacent in size due to the absorption of interstitials by smaller clusters and the emission of interstitials by larger clusters, $x_{C_I}$ and $x_{Si_I}$, respectively, are ratios of $C_I$ and $Si_I$ in a cluster. Here, they are equal to 0.5 because of our stoichiometric assumption. The factor of 2 on the right hand sides of Equation *3* and Equation *4* accounts for the case when two monomers have been joined together to form a cluster of size two. $P_{C_I}$ represents the production rate of $C_I$ by irradiation, $Q_{C_I}$ represents the rate of change of the atomic fraction of a given PD (here $C_I$) due to reactions among PDs. The net flux between adjacent clusters can be written as:

$$J_{n \to n+1} = \beta_{n,n+1} C_n - \alpha_{n+1} C_{n+1} \qquad \text{Equation 6}$$

where, $\beta_{n,n+1}$ is the absorption rate of an interstitial by cluster of size $n$. Clusters in irradiated 3C-SiC are assumed to be planar. Planar geometry has been directly observed for large clusters [4, 21, 22]. This assumption is also adequate for small clusters, which are neither spherical nor planar[23]. The absorption rate is then defined as [12, 14]

$$\beta_{n,n+1} = 2\pi(r_n + r_1)D_{eff}^d/\Omega \qquad \text{Equation 7}$$

Here, $\Omega$ is the average atomic volume in 3C-SiC and $r_n$ is the radius of the planar clusters with burgers vector $b = <111>$. $D_{eff}^d$ is the effective diffusion coefficient of monomers that can be absorbed by a cluster. As there are two kinds of mobile defects in our system, namely $C_I$ and $Si_I$, we follow the definition of the effective diffusion coefficient in multi-component alloys derived by Slezov [24]:

$$D_{eff}^d = \left(\frac{x_{C_I}^2}{C_{C_I}D_{C_I}} + \frac{x_{Si_I}^2}{C_{Si_I}D_{Si_I}}\right)^{-1} \qquad \text{Equation 8}$$

$D_{C_I}$ and $D_{Si_I}$ are the diffusion coefficients of $C_I$ and $Si_I$, respectively. $C_{C_I}$ and $C_{Si_I}$ are concentrations of of $C_I$ and $Si_I$, respectively.

Interstitials in SiC are quite mobile and they can rapidly aggregate into clusters, therefore it is reasonable to assume that the concentration of interstitials in the 3C-SiC matrix is dilute. With this assumption, the real time equilibrium cluster distribution can be derived from the classical nucleation theory as

$$\overline{C_n} = exp\left(-\frac{\Delta G_n}{kT}\right), \qquad \text{Equation 9}$$

where $\Delta G_n$ is the change in the Gibbs free energy of the system when a cluster of size *n* is formed from *n* monomers. $\Delta G_n$ is calculated as

$$\Delta G_n = E_f^{SiC}(n) - x_{C_I} \cdot n \cdot \left[E_f^{C_I} + kT\ln C_{C_I}\right] - x_{Si_I} \cdot n \cdot \left[E_f^{Si_I} + kT\ln C_{Si}\right]. \qquad \text{Equation 10}$$

Here, $E_f^{SiC}(n)$ is the formation energy of cluster of size *n*, which will be discussed later. $E_f^{C_I}$ and $E_f^{Si_I}$ are formation energies of $C_I$ and $Si_I$, respectively, which have been calculated with DFT[19].

The emission rate $\alpha_{n+1}$ of an interstitial from a cluster can be derived based on two assumptions. One is the constrained equilibrium assumption, with which all fluxes $J_{n \to n+1}$ among adjacent clusters equal zero at equilibrium. Secondly, we assume that the emission rate is an intrinsic property of clusters, so it can be derived for a solution with any nominal concentrations of interstitials. With these two assumptions and Equation 9, the emission rate can be determined as follows:

$$\alpha_{n+1} = \beta_{n,n+1}\frac{\overline{C_n}}{\overline{C_{n+1}}} = \beta_{n,n+1} \cdot exp\left(-\frac{\Delta G_n - \Delta G_{n+1}}{kT}\right) \qquad \text{Equation 11}$$

The PDs production rate is given by the following expression

$$P_{PD} = \Gamma \eta \xi_{PD} \qquad \text{Equation 12}$$

where $\Gamma$ is the dose rate in dpa/s. Based on The Stopping and Range of Ions in Matter (SRIM) [25] calculations, $\Gamma = 2.0 \times 10^{-3}$ dpa/s. $\eta$ is the cascade efficiency, which describes the efficiency of successfully displacing one atom from its original lattice site. $\xi_{PD}$ is the fraction of a certain kind of PD produced during irradiation. We adopt the values of $\xi_{PD}$ determined through molecular dynamics (MD) simulations of displacement cascades by Swaminathan *et al*. [17].

One thing should be noted is that the model introduced above is the basic model used in the first part of the paper and this model is based on known properties of defects and well-established phenomena. Later on we will introduce modifications to the above CD equations, based on other physical phenomena that may be potentially taking place in irradiated SiC and we will illustrate the effect of such phenomena.

## 3 Results
### 3.1 TEM and STEM analysis

**Figure *2*** (a) shows a low angle annular dark field (LAADF) STEM image of Kr-irradiated 3C-SiC with the viewing direction along the <110> axis. Compared to high-angle annular dark field (HAADF) STEM images, LAADF images emphasize strain contrast, making small BSDs more visible [26]. The bright spots are the Si atom columns, as shown in the inset. C columns are not resolved. BSDs are dark patches in the image arising from lattice distortion. Some BSDs appear to be spherical. Other BSDs are elongated within the {111} or {001} planes, as indicated by red circles. Control images (not shown here) from unirradiated regions of the same sample show no black patch contrast. High-angle annular dark-field images, which emphasize Z-contrast over strain from irradiated regions, also do not show contrast from BSDs. **Figure *2*** (b) shows a bright field TEM image of irradiated 3C-SiC. BSDs in this image show dark contrast and take mostly circular or slightly oval shape, as marked by a circle. Other types of defects, such as loops, voids and a dislocation network were not found in this sample.

We used the conventional particle analysis routines as implemented in DigitalMicrograph software to identify BSDs in BF TEM images, such as the one shown in **Figure *2*** (b). Broadly speaking, this method defines an intensity threshold, and then looks for clusters of pixels with intensities below the threshold, and these clusters are considered to be defects. We analyzed three images like the one shown in **Figure *2*** (b) and identified 672 total defects.

Unfortunately, a single threshold particle analysis is ineffective for high-resolution LAADF STEM images, because of the bright atomic columns present in the images. Therefore, we developed an identification scheme described below that uses multiple, local intensity thresholds to find BSDs against the varying background contrast of the atomic columns and implemented it in MATLAB. **Figure 3** illustrates the steps in the algorithm on an example image. Once the defects have been correctly identified in a binary image, we can use conventional particle analysis techniques[27] can be used to determine their size, density, *etc*.

**Figure 3** (a) is a gray-scale STEM image showing BSDs as dark contrast superimposed on the bright contrast of the atoms. The squares identify regions that contain a BSD, and the circles indicate BSD-free regions. The most obvious feature of the BSDs is that they reduce the intensity between the atomic columns in the image, so the first step in the analysis is to identify the positions of all the atomic columns and remove them from further consideration. This procedure results in **Figure 3** (b), in which the small white regions represent atomic columns that were removed from further analysis. The next step is to identify all of the pixels in the image that could possibly participate in a BSD using a relatively high global intensity threshold, which is 128 in this image. **Figure 3** (c) shows the result as a binary image: all the black pixels could be part of a BSD and the white pixels are excluded. However, **Figure 3** (c) does not show discrete defects like the ones shown in the squares in **Figure 3** (a). Because of TEM sample thickness variation, no global intensity threshold could effectively detect small BSDs. The next step in the analysis is therefore to apply a local intensity threshold. We divide the image into sub-regions such as the squares in **Figure 3** (a). We then apply a threshold equal to the mean intensity within each region, resulting in the image shown in **Figure 3** (d). Here, black regions are more compact and more closely resemble the BSDs in **Figure 3** (a).

The circles in **Figure 3** (d), where there are no BSDs, show some examples where the local threshold has found single or a few dark pixels and ascribe them to BSDs. We can remove these artifacts by requiring BSDs to exceed a minimum size. **Figure 3** (e) shows this minimum size, approximately one atom spacing, superimposed on a small portion of the original image (**Figure 3** (a)) and the local threshold image (**Figure 3** (e)). Any region of the image smaller than this could not accommodate more than one interstitial, and probably contains none, so we remove it as a candidate BSD. **Figure 3** (f) shows the binary mask in **Figure 3** (e) after removing too-small regions.

The last step is to add the atomic column pixels back into the analysis to create a binary mask identifying all the pixels inside the BSDs as black and all the other pixels as white, suitable for particle analysis. All the atomic columns from **Figure 3** (b) that are surrounded entirely by BSD regions in **Figure 3** (f) are considered to be part of the BSDs, and all the other columns are not. The result is shown in **Figure 3** (g) as a binary mask superimposed on the original image from **Figure 3** (a). In **Figure 3** (g), all pixels shown in black are identified as belonging to BSDs. All the other pixels are copied from **Figure 3** (a) unchanged for display purposes. The end result is an image of BSDs that is amenable to conventional particle analysis, large enough to be physically reasonable, and a good match to the human eye identification of BSDs in **Figure 3** (a). This automated procedure was applied to 21 images, resulting in identification and measurement of 543 total BSDs.

**Figure 4** shows size distribution of clusters in irradiated 3C-SiC determined from LAADF STEM and BF TEM imaging. Area densities of defects are converted to volume densities using the measured thickness of the TEM sample for each image. Each histogram bin in **Figure 4** is reported with $\sqrt{N}$ error bars, where $N$ is the number of defects counted in each bin. This error bar dominates the other experimental errors arising from, for example, estimate of the sample thickness. The integrated size distribution uses the more reliable STEM results for clusters size smaller than 1 nm in diameter, and the average of STEM and TEM results to achieve better statistics for cluster size larger than 1 nm in diameter.

The defect diameters range from about 0.4 nm to 8.3 nm. The TEM and STEM data are in good agreement for diameters larger than 1.0 nm, although the TEM data are more statistically reliable due to the larger number of counted defects. Below 1.0 nm, STEM detects a much larger concentration of defects and the defect concentration from STEM is approximately 1.5 times higher than from the TEM data. The reason for this discrepancy lies in the difference in resolution limits between TEM and STEM. As reported by Zhou *et al.* [28], the resolution of TEM is limited to 1 nm due to divergence of Bragg diffracted beams and the failure of the column approximation. The resolution of STEM used in this study is as low as 0.8 Å. However, the sensitivity of STEM to defect clusters is determined not just by the instrument resolution, but also by the cluster structure, TEM sample thickness, and imaging conditions. One can estimate this detection sensitivity by simulating STEM image of a well-organized cluster, such as a dislocation loop with a well-defined strain field. To do that, we have used the multislice technique and simulated STEM image of an interstitial Frank loop in 4H-SiC; this loop had a diameter of 1.4 nm and a 2.4 nm thickness. We were able to detect strain as low as 2%, giving a -1% contrast in low LAADF STEM image. This analysis shows that STEM is a reasonable approach to imaging sub-nanometer defect clusters with a significant lattice distortion. In the STEM data, we found ~74% of the defects to be smaller than 1 nm, as compared to ~47% in the TEM data. The total defect density, integrating over all cluster diameters, in STEM is $(50 \pm 0.11) \times 10^{24}$ m$^{-3}$ and $(1.19 \pm 0.05) \times 10^{24}$ m$^{-3}$ by TEM. We believe that the higher strain sensitivity of the LAADF STEM images compared BF TEM results in this increased ability to detect small diameter defects.

The impact of the high density of small defects (< 1 nm) observed in STEM on swelling is discussed in Section 4. The calculations in the following sections are based on the integrated size distribution shown in **Figure *4*.**

## 3.2 Cluster dynamics model with known physical parameters for SiC
### 3.2.1 Parameters of the model

Since the CD method is quite sensitive to the specific values of parameters and many of the parameters are often unknown, it is common to optimize some of these parameters to fit experimental data [29, 30]. Here, we begin with the known parameters and known physics related to defect behavior in irradiated SiC and we fit the parameters to experimental data where necessary.

Formation energies of PDs and defect clusters constitute one of the important sets of parameters that enter CD model. Formation energies of PDs in SiC are generally known and can be found for instance in Refs. [17-19, 31-33] (see also **Table *1***). Here we use values reported by Shrader *et al*. [19], which were calculated using DFT. For defect clusters we considered formation energies calculated by Jiang *et al*. [31] using DFT (open square symbols in **Figure *5***) and by Watanabe *et al*. [34] using classical empirical potentials (open diamond and open circles in **Figure *5***). While DFT is a more accurate method, it is limited to small cluster sizes (smaller than 6 interstitials in this case). Another difference between Jiang *et al*.'s and Watanabe *et al*.'s approaches is that the latter study assumed clusters to be disc-shaped, whereas DFT calculations by Jiang *et al*. found that this assumption is not necessarily true for small clusters. Jiang *et al* found that small clusters are neither spherical nor planar. Formation energies of clusters as a function of their size calculated by different approaches are summarized in **Figure *5***. Although discrepancies exist among these calculations, in this paper we treat them as a guide for a range of possible values. Interestingly, the energies predicted by DFT and empirical potentials seem to be mainly shifted with respect to each other, and here we use a number of schemes to combine these two approaches. They include a *switching function* method (blue lines) and a *shift* method (red line), both illustrated in **Figure 5**. The switching function has the following analytical form

$$S(n) = \begin{cases} 1 & n = 1 \\ \dfrac{(7^2 - n^2)^2(7^2 + 2n^2 - 3 \cdot 1^2)}{(7^2 - 1^2)} & 1 < n < 7 \\ 0 & n = 7 \end{cases} \qquad \text{Equation 13}$$

in the regime where Jiang *et al.*'s and Watanabe *et al.*'s calculations overlap. Here, *n* is the number of interstitials in the cluster. The resulting formation energy of a cluster can therefore be written as

$$E_f^{switching}(n) = E_f^{\text{DFT}}(n) \cdot S(n) + E_f^{\text{MD}}(n) \cdot (1 - S(n)) \qquad \text{Equation 14}$$

where $E_f^{\text{DFT}}$ means the formation energy calculated by Jiang *et al.* with DFT, and $E_f^{\text{MD}}$ represents the formation energy calculated by Watanabe *et al.* using empirical potentials.

In the shift method, formation energies for clusters smaller than seven are taken from Jiang *et al.* [31]. For larger clusters, Watanabe *et al.*'s results [34] were shifted to match the DFT value for *n* = 6 and to maintain a smooth trend as a function of size.

Migration energies for PDs are taken from Refs. [8, 35]. All reactions and reaction barriers adopted in our CD model are summarized in **Table 2**. Initially, we assume the defect clusters to be immobile at the experimental temperatures, but later in this paper we will consider the effect of cluster mobility on the cluster size distribution.

Cascade efficiency, $\eta$ depends on the incident atom (i.e., its species, energy, and velocity direction), as well as on the matrix material composition and structure [36]. In the last thirty years various values for η have been reported from experiments [37-43] and simulations [44-49]. This range of values is shown in **Figure 6** (including data for SiC) and in our model we will treat η as a parameter fitted within a range of the previously observed values (0.001 to 1).

The total simulation time in our CD model is 120 s, which is the same as the duration of the irradiation experiments. In order to validate our simulations, all results are reported at this simulation time and compared with experiments results.

### 3.2.2 Predicted cluster size distribution

There are two sets of parameters that we allow to vary in our simulations: cluster formation energies and cascade efficiencies (See Sec.3.2.1). In order to optimize their values, we minimize the define root mean square deviation (RMSD) between the model predictions and experiments, which is defined as

$$\text{RMSD} = \left( \frac{1}{5} \sum_{i=1}^{5} (C_i^{\text{sim}} - C_i^{\text{exp}})^2 \right)^{0.5} \qquad \text{Equation 15}$$

$C_i^{\text{sim}}$ and $C_i^{\text{exp}}$, respectively, represent simulation and experimental concentrations of clusters belonging to the $i^{\text{th}}$ bin in the histogram representing the number density vs the cluster diameter. Only clusters with diameters equal to or smaller than 2.5 nm (corresponding to the first 5 columns in *Figure 7*) are taken into account when calculating RMSD as other values are too small to allow reliable statistics.

Interestingly, the results are not strongly dependent on the specific functional form (switching method or shift method in **Figure 5**) of the cluster formation energy vs the cluster size for the entire range of efficiencies considered here. This lack of dependency can be understood to arise from the fact that for both of these functions formation energies of isolated interstitials are quite high as compared to the average formation energies of interstitials in clusters. As a consequence, processes involving isolated interstitials coming together to form clusters or joining existing clusters are energetically very favorable, and the dissociation rate of clusters is negligible as compared to the absorption rate. The entire evolution is thus controlled by the cluster absorption rate and the effects arising from different cluster formation energy functions are negligible (e.g., the RMSDs varies by less than 1%). As the cluster radius $r_n$ and the atomic volume $\Omega$ are fixed, the evolution is fundamentally controlled by the mobility of interstitials as shown in Equation 7. Since our results are weakly dependent on the specific function representing the cluster formation energy, in the remainder of this paper we choose one of them (the switching function) to present the results of our model.

We have also explored the effect of cascade efficiency on the cluster size distribution (**Figure 7** (a)). The smallest RMSD, $6.19888 \times 10^{23}$ 1/m$^3$, is found for $\eta = 0.004$, although RMSD values are very large and essentially identical for all values of $\eta$. Specifically, RMSD varies from $6.1989 \times 10^{23}$ to $6.199 \times 10^{23}$ 1/m$^3$ when $\eta$ is in the range from 0.001 and 1.0. The main contribution to the RMSD value comes from the fact that experimental cluster densities are 2 to 3 orders of magnitude higher than the ones predicted in simulations. The order of magnitude in the predicted number densities does not change with the cascade efficiency within the range of $\eta$ values reported in literature and shown in **Figure 7** (a). In **Figure 7** (b) we plot the cluster size distribution measured experimentally and predicted by simulation using the lowest value of the cascade efficiency reported in literature $\eta = 0.006$. Not only is the order of magnitude of the number density is different but also the peak of the distribution predicted by simulation is shifted to the right with respect to the experimental peak. Increasing the value of $\eta$ shifts the simulation peak further away from the experimental peak (**Figure 7** (a)). This analysis shows that with the current version of the CD model (which includes known physics of defects in irradiated SiC), it is not possible to match the experimental cluster size distribution.

The above results provide guidance into what physical phenomena might be missing in the current understanding of irradiation phenomena in SiC. First of all, one should note that for all values of $\eta$ that can be physically justified (based on **Figure 6**), the majority of cluster sizes vary from 2.0 nm to 6.0 nm (*Figure 7* (a)). As mentioned above, these values are larger than the peak in the cluster size density measured in TEM and STEM experiments (see *Figure 7* (b)). These results indicate that the cluster nucleation phase is too short and the growth of clusters is too fast in simulations as compared to experiments. It might be intuitive to think that one could correct this issue by introducing more PDs in the model (through increasing cascade efficiency), because more PDs would lead to faster nucleation and therefore to an increase in the number of small clusters. However, PDs also provide the driving force for cluster growth, and as shown in *Figure 7* (a) the net result of increasing $\eta$ is to shift the cluster size distribution predicted in the CD model even further to the right (toward larger clusters). In subsequent sections we will explore what potential physical phenomena (that are plausible but perhaps not yet established for SiC) can alter the predicted cluster size distribution and bring it into a better agreement with experiment.

## 3.3 Effects of potential physical phenomena in irradiated SiC on predictions of the CD model

### 3.3.1 Effect of intra-cascade cluster production

It is possible that some of the clusters form directly in the cascade, instead of arising only from interaction of PDs in the cascade settling stage. This phenomenon was captured by TEM [50] and found to be necessary in CD models of irradiated tungsten, molybdenum, and iron in order to reproduce experimental cluster size distributions [13, 30, 51, 52]. Since currently there is no model for predicting the size distribution of clusters produced in the cascade as a function of the PKA energy or ion energy in SiC, we adopted the functional form for such distribution from Xu *et al*. [13], which was first developed based on studies in iron and then applied to Mo. The experiments in Ref. [13] were conducted with the same chemical identity and energy of the incident ion as in our study, and the intra-cascade cluster production function $\xi_n$ has been integrated over a range of PKA energies. We used the same functional form but scaled the distribution function to make sure the sum of interstitials in all intra-cascade interstitial clusters matches the total production of C and Si interstitials in SiC, as predicted using MD simulations by Swaminathan *et al*. [53]. Including intra-cascade cluster production in the CD model changes the evolution function described by Equation *2* to the following form

$$\frac{dC_n}{dt} = P_n + J_{n-1 \to n} - J_{n \to n+1} \qquad \text{Equation 16}$$

Here the production term $P_n$ can be expressed as

$$P_n = \Gamma \eta \xi_n, \qquad \text{Equation 17}$$

where $\Gamma$ is the dose rate in dpa/s and $\eta$ is the cascade efficiency, which has been discussed earlier. $\xi_n$ is the intra-cascade cluster production function discussed above, which represents the number of clusters of size *n* generated per one atomic displacement.

*Figure 8* (a) shows the results of simulations that include the intra-cascade cluster production for cascade efficiencies that vary from 0.001 to 1.0. Most of the clusters in simulations are smaller than 1 nm (irrespectively of the specific value of η) and this time the model cannot reproduce the non-negligible concentrations of larger clusters observed in experiments. The reason for a suppressed growth of clusters (and therefore for the lack of larger clusters in simulations) is that intra-cascade clusters act as nucleation sites and they quickly consume most of the PDs in the matrix. As clusters in SiC are significantly more stable than PDs (see **Figure *5***), the dissociation rate of clusters is negligible. Depletion of PDs from the matrix reduced the driving force for cluster growth.

Since the specific function for production of intra-cascade clusters can affect the results, we have explored the effect of changing this function on the predicted cluster size distribution. Specifically, we assume that the number of interstitials in intra-cascade clusters obey a power law $A \times n^{-B}$. This relationship is qualitatively consistent with cluster size distribution in irradiated 3C-SiC [54] and with previously reported results for W, Fe, Be, Zr [55-58] and Mo [59]. Here, we vary B from 0.0 to 2.5, which covers the range of intra-cascade cluster production functions from cascade simulations in Refs. [44, 55-58, 60, 61]. We also vary the largest intra-cascade cluster size, $n_{max}^{Intra-cascade}$ between 4 (which is the same as in MD simulations by Gao *et al.*[44]), and 21 (which is chosen to be one interstitial larger than the largest intra-cascade cluster size in Xu *et al.* [13]). In each case we calculate A by ensuring that the total number of interstitials in all intra-cascade interstitial clusters matches the total production of C and Si interstitials in SiC, as predicted by MD simulations of Swaminathan *et al*. [53]. Results of our CD simulations for the limiting values of M and $n_{max}^{Intra-cascade}$ are shown in **Figure 8** (b). It can be seen that the vast majority of clusters observed in simulations have a size of less than 1nm and the cluster density is around $10^{23}$ 1/m$^3$. Introducing intra-cascade cluster production therefore can bring down the peak cluster size from 2 nm – 6 nm (in **Figure *7***) to 0 nm – 1 nm (in **Figure *8***). However, the number density of the simulation results has been changed from 1 magnitude smaller than the experimental result, to 1-4 orders of magnitude higher than the experimental data.

These results show that introducing intra-cascade cluster production alone over-corrected the discrepancy in the density of clusters between experiments and simulations observed in ***Figure 7***. It also led to too many small clusters being present in the matrix and correspondingly to the lack of experimentally observed larger clusters.

### 3.3.2 Effect of cluster migration

Except for Si di-interstitials studied in Ref. [62], defect clusters in SiC have been generally assumed to be immobile [3] due to their high migration barriers. For instance, Jiang *et al.* [63] used accelerated molecular simulations to show that migration barrier for C tri-interstitial cluster is ~4.12 eV, which means that this cluster will be immobile on experimental time scales up to temperatures of 1,200K (at these temperatures the cluster performs ~ 1 hop/day). However, there is recent experimental evidence that under some irradiation conditions clusters in SiC can undergo radiation-induced diffusion due to ballistic collision of the incoming particles with cluster atoms[26]. This phenomenon has been discovered in SiC only recently and the effective migration barriers of clusters during irradiation of SiC are unknown. However, a number of different approaches to estimate diffusion coefficients of clusters have been put forth in the literature on metallic systems [13, 14, 30, 52, 64, 65]. All of them share the common feature that the cluster diffusion coefficients decrease as the cluster size increase. Here, we explore the effect of assuming clusters are mobile and we approximate the energy barrier, $E_m^n$, of a stoichiometric cluster of size *n*, using the following function:

$$E_m^n = E_m^{Si_I} + F \times (n-1) \qquad \text{Equation 18}$$

In the above equation we used the migration barrier of Si interstitial $E_m^{Si_I}$ because it is higher than the migration energy of C interstitial. *F* is a positive constant that can be optimized based on the RMSD calculation. Diffusion pre-factors of clusters are taken as equal to the pre-factor for diffusion of Si interstitial.

In ***Figure 9*** (a) we show the calculated RMSD values for clusters smaller than 2.5 nm (as defined in Equation *15*) for a range of cascade efficiencies η from 0.001 to 1.0 and for a range of *F* parameter values from 0.05 to 1 (*F* is defined through Equation 18). Similarly as we found using the basic CD model (see Section 3.2), the optimized cascade efficiency is small ($\eta = 0.002$), and RMSD values are large across most combinations of $\eta$ and $F$.

The minimum RMSD for the entire range of parameters shown in **Figure 9** (a) is $1.667 \times 10^{23}$ $1/m^{-3}$ and it corresponds to the cascade efficiency $\eta$ of 0.002 and the migration barrier common difference $F$ of 0.75 eV. Using the above values of $\eta$ and $F$, we simulated the cluster size distribution and the results are plotted in **Figure 9** (b). The agreement between simulations and experiments is good for clusters smaller than 1 nm, but clusters larger than this value are missing from simulations. If we want to optimize the agreement for larger clusters, we can exclude clusters smaller than 1 nm and in that case the smallest RMSD is found to be $2.998 \times 10^{22}$ $1/m^3$. The corresponding values of $\eta$ and $F$ are 0.007 and 0.2 eV, respectively, and the predicted cluster size distribution is compared to experiments in **Figure 9** (b). **Figure 9** (b) reveals that decreasing the migration barrier (i.e., increasing $F$) moves the cluster distribution peak towards larger sizes. This trend can be understood by the fact that a lower migration barrier increases the mobility of clusters. The clusters can run into each other and coalesce, which process increases the rate of formation of large clusters. The observations that with clusters allowed to migrate it is possible to reach the same order of magnitude in cluster size distribution between simulations and experiments and that by changing migration of clusters (controlled by $F$) we can match quantitatively either the small or the large clusters in the distribution suggests that perhaps clusters of the same size may have different migration barriers. One possible reason for this behavior could be morphology of the cluster, which is discussed in the next section.

### 3.3.3 Effect of cluster morphology

Although larger black spot defects and dislocation loops (as observed in TEM) have been shown to be mainly planar in irradiated SiC [4, 21, 22], some authors pointed out that very small clusters can be non-planar in SiC [4, 23]. This assertion is supported by atomistic simulations of Jiang *et al*. [31, 63]. Our inspection of STEM images, such as **Figure 2** (a), reveals that on the average approximately 59% of clusters smaller than 1 nm lie within the {111} planes with a deviation of $\pm 30°$, and among clusters larger than 1 nm about 82% of cluster lie within the {111} planes with $\pm 30°$ angle error bar. These results show that during irradiation and post-irradiation defect evolution, there is a non-negligible fraction of clusters that do not lie on the {111} planes. These out-of-plane clusters likely have different properties than the in-plane clusters. Since larger clusters and loops are observed mainly to be in-plane, we assume here that growth of the out-of-plane clusters is energetically unfavorable because of the high energy cost associated with breaking bonds during the geometry change from out-of-plane to in-plane. Since these clusters are still observed in our samples, we also assume that they do not move easily, dissolve or transition to in-plane configurations. Finally, as pointed out in Sec. 2.2 out-of-plane clusters cannot annihilate with vacancies at the temperature of our experiments because the vacancies are immobile. Again, the functional form for production of out-of-plane clusters is not available. Here, we assume it obeys a power law function $\xi_n^{out-of-plane} = A \times n^{-B}$ with $A$ and $B$ being adjustable parameters in our model. $A$ varies from $1 \times 10^{-4}$ to $1 \times 10^{-6}$ and $B$ varies from 0.6 to 1.2 to make sure $\xi_n^{out-of-plane}$ is smaller than the total number of clusters derived in Section 3.3.1 based on results of Xu *et al*. [13]. By analyzing RMSDs from simulations with various $A$ and $B$ values, we found $\xi_n^{out-of-plane} = 6 \times 10^{-5} \times n^{-1.2}$, $\eta = 0.007$, and $F = 0.2$ eV to give the best match between simulations and experiments (see **Figure *10***). The total number of interstitials in out-of-plane intra-cascade clusters is only 0.1% of all the interstitials produced by a cascade. However, due to their lack of mobility, and the resultant lack of coalescence and unfavorable growth, out-of-plane clusters quickly accumulate during irradiation. In the end, they constitute the majority of clusters smaller than 1nm.

As mentioned above, the optimized cascade efficiency is found to be 0.007, which is relatively small as compared with recent MD simulations in SiC. It is possible that MD overestimates cascade efficiency because it does not account for the energy losses through inelastic interactions, such as electron excitation and ionization. Although Stoller *et al.* [66] separated the elastic energy loss from inelastic energy loss with the procedure described by Norgett *et al.* [67] many years ago, the coupled effects of energy dissipation through elastic and inelastic interactions on defect production was very seldom discussed. These effects could be competing, additive, or synergetic. Recent research conducted by Zhang *et al.* [68] revealed that inelastic energy loss leads to local heating via electron-phonon coupling, and may lead to damage recovery instead of damage (defect) production in SiC. Furthermore, Zhang *et al.* [68] found evidence that radiation damage caused by 900 keV Si ion can be fully healed as the radiation dose increases to the regime where the electronic stopping power is the dominant energy loos mechanism. In the current state of knowledge it is not clear whether the same arguments could be applied to the irradiation conditions in our experiments. It is also possible this discrepancy is caused by the uncertainties of energy barriers for point defect recombination [8]. For example, if the barriers in the CD model are higher than the true values, the model would predict more interstitials being left in the matrix. This in turn would accelerate cluster growth rate and would force a low value of cascade efficiency during optimization of the model to decrease interstitial supply from cascades.

To summarize, we find that direct simulation with the current understanding of fundamental radiation effects in SiC creates far too small cluster densities. Adding in intra-cascade cluster production can yield densities in the correct range, but creates too many small clusters. Further adding cluster migration can then reproduce the larger (~>1nm) portion of the particle size distribution but the agglomeration leaves too few small clusters. Finally, adding in cluster production of clusters that are inactive for diffusion or growth can match the complete measured particle size distribution. Although inclusion of intra-cascade cluster production, mobile clusters, and out-of-plane cluster production is to some extent speculative, there is evidence (as discussed above) that these phenomena might be present in irradiated SiC. In addition our simulations demonstrate that with the current understanding of fundamental radiation effects in SiC, it is not possible to reproduce experimental cluster distribution in this material. Our results provide new insights into what other defect-related phenomena may exist and what would be their effect on defect evolution in irradiated SiC.

## 4   Swelling

Having proposed a CD model of cluster evolution in irradiated SiC, it is interesting to ask what is the amount of swelling predicted by this model and how it compares to values reported for similar irradiation conditions. Here we include contributions to swelling from both defect clusters and from isolated PDs that are present in the lattice.

The number of PDs present in irradiated samples can be estimated as follows. As both carbon and silicon interstitials diffuse very fast and have a strong tendency to cluster, there are few isolated interstitials left in matrix. On the other hand, once an interstitial is trapped in a cluster, it is difficult for this interstitial to be emitted or react with isolated vacancies or antisites (since these the latter two defects have relatively high migration barriers and do no diffuse easily – see **Table 2**). Our CD model shows the ratios between the number of interstitials trapped in clusters and the number of isolated interstitials, vacancies, and antisites reach approximately constant values as the simulation progresses. The CD model, which is optimized to reproduce clusters size distribution for clusters with diameters $D < 2.5$ nm, can be used to determine the number of isolated interstitials, vacancies, and antisites. The number of interstitials trapped in clusters with $D < 2.5$ nm can be calculated based on the experimental cluster size distribution by multiplying the estimated cluster area by the Burgers vector, and then dividing by the atomic volume of 3C-SiC. The ratios of interstitials trapped in clusters over isolated interstitials, vacancies, and antisites calculated using this method are listed in **Table 3**. We then calculate the total number of interstitials trapped as clusters and in this calculation we count clusters of all sizes observed in TEM and STEM, including those with $D > 2.5$ nm. The total number of trapped interstitials is then multiplied by the ratios from **Table 3** to get the total number of isolated points defects that are present in the samples (it is not possible to be observed these PDs directly).

Swelling caused by the presence of clusters can be calculated by multiplying their areas by their Burgers vector. Contributions to swelling from isolated PDs can be derived by multiplying their concentrations by their formation volumes, which have been determined in Ref. [32] and are reported in **Table 3**.

Using the above approach, the total swelling is calculated to be 0.155% in our sample, which was irradiated with 1 MeV Kr$^+$ at temperature 1073 K to a fluence of $3\times10^{14}$ Kr/cm$^2$. This swelling was measured at the location in the sample that corresponds to the dpa of 0.24. Our estimated swelling is on the same magnitudes the results of Price *et al*. [69], who irradiated 3C-SiC at 1053 K with fast neutrons (E > 0.18 MeV) to $6\times10^{21}$ n/cm$^2$, and the measured the swelling was 0.4±0.1%.

In our study we find that 38.13% of the total swelling comes from isolated PDs, 10.77% of the total swelling comes from BSDs or smaller clusters ($D < 1.0$ nm) where TEM underestimates the cluster concentration, and 51.10% of the total swelling is caused by larger clusters or dislocation loops ($D > 1.0$ nm). This analysis is consistent with the observation that volume expansion calculated based on the concentration of defects loops visible in traditional TEM accounts only for one third of the total swelling in SiC [7]. The mismatch between TEM based calculation and measured swelling is largely due to the small BSDs and isolated PDs that cannot be easily characterized by TEM. In addition, from the 38.13% of swelling that is due to PDs, interstitials contribute only less than 0.01% of swelling (most of the interstitials are found in clusters), while vacancy and antisites contribute 33.35% and 66.65%, respectively.

## 5  Conclusion

The microstructure of 3C-SiC under 1 MeV krypton ion irradiation at 800 ℃ was studied using both TEM and STEM experiments and CD simulations. We found that there are very small clusters ($D < 1$ nm) presented in irradiated SiC that are missing in traditional TEM analysis due to resolution limitations. A complete characterization of defect clusters in SiC should therefore involve high resolution STEM. CD model based on the known physics of defects in SiC, such as PDs productions, formation and migration energies, and reported cascade efficiencies is unable to reproduce the experimentally determined cluster size distribution. By comparing CD simulations with experiment, we have proposed other phenomena to be present in irradiated SiC, which could have a significant impact on cluster size distribution. These phenomena include intra-cascade cluster production, irradiation-induced mobility of clusters, and clusters with different morphologies. We have shown that by including the above physics into the model it is possible to obtain agreement with experiments for the irradiation conditions reported in this paper. It will be important to investigate these phenomena in detail in order to develop a physics-based model that is transferable to other irradiation conditions. A swelling model has been developed based on our CD model. It provides an explanation for the mismatch between swelling measured experimentally and swelling estimated based on loops observed in traditional TEM examinations. This mismatch is due to the presence of very small clusters and isolated PDs, which both have a non-negligible contribution to the total swelling.

# 6 Acknowledgement

The author gratefully acknowledge financial support from Department of Energy, Office of Nuclear Engineering under Nuclear Engineering University Program (NEUP) Grant DE-NE0008418. We also acknowledge helpful discussions with Dr. Yutai Katoh and Dr. Takaaki Koyanagi from ORNL and Dr. Mingjie Zheng, Dr. Chao Jiang, and Dr. Xing Wang from UW-Madison.

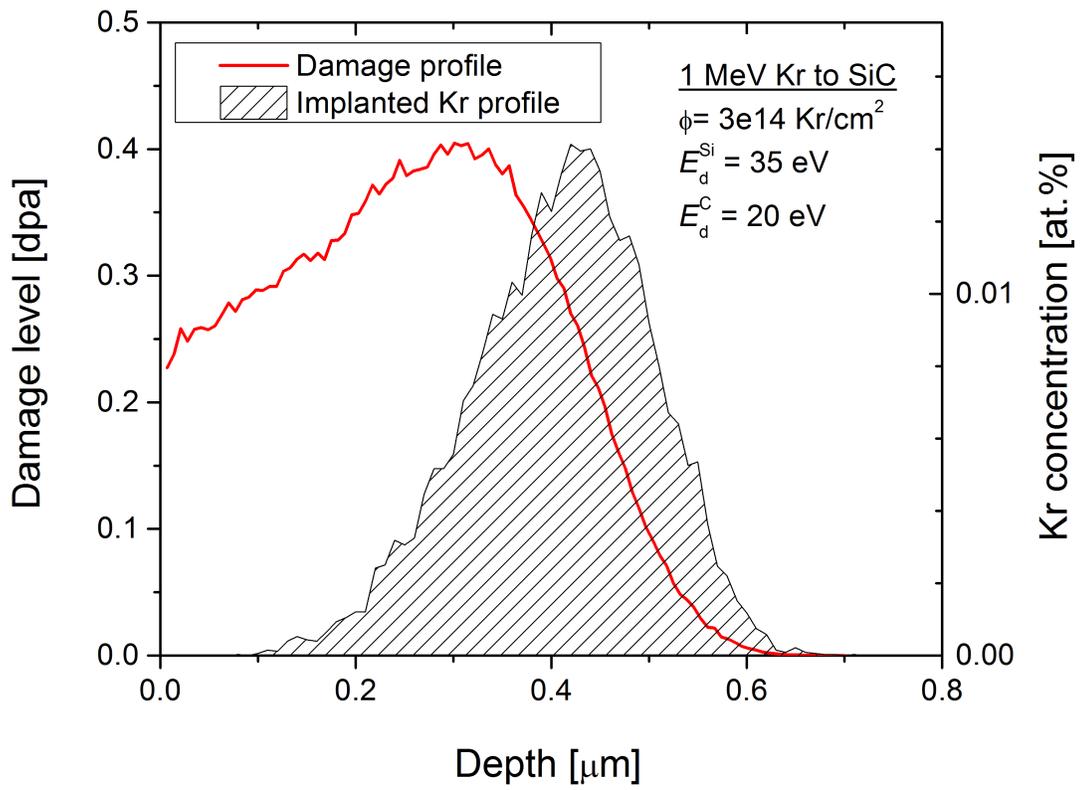

**Figure 1**: (Color online) Damage and Kr-implantation profile in 1 MeV Kr irradiation of SiC

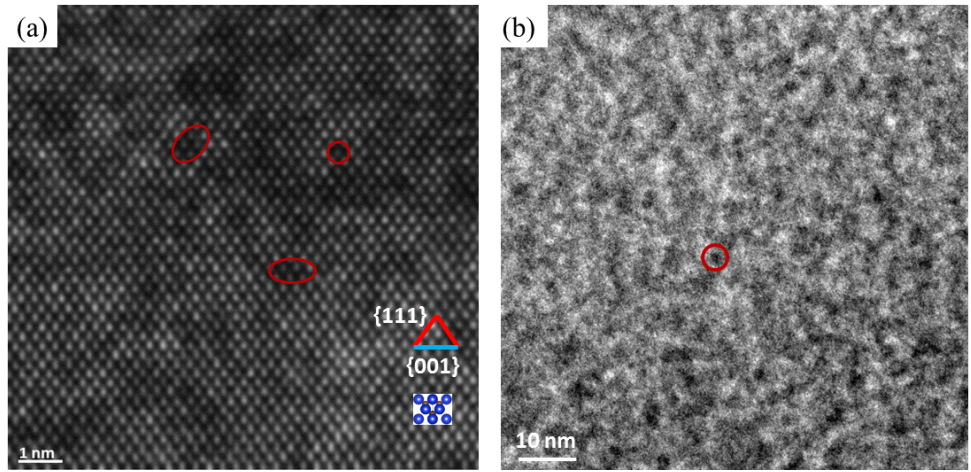

**Figure 2:** (a) LAADF-STEM image of BSDs in 3C-SiC. The viewing direction is along the zone axis <110>. The projection of two {111} and one {001} planes are indicated by the red and blue lines, respectively. Inset shows the SiC crystal structure with Si atoms in blue and C in red. (b) TEM bright field image of 3C-SiC along the zone axis <110>. BSD examples are indicated by red circles.

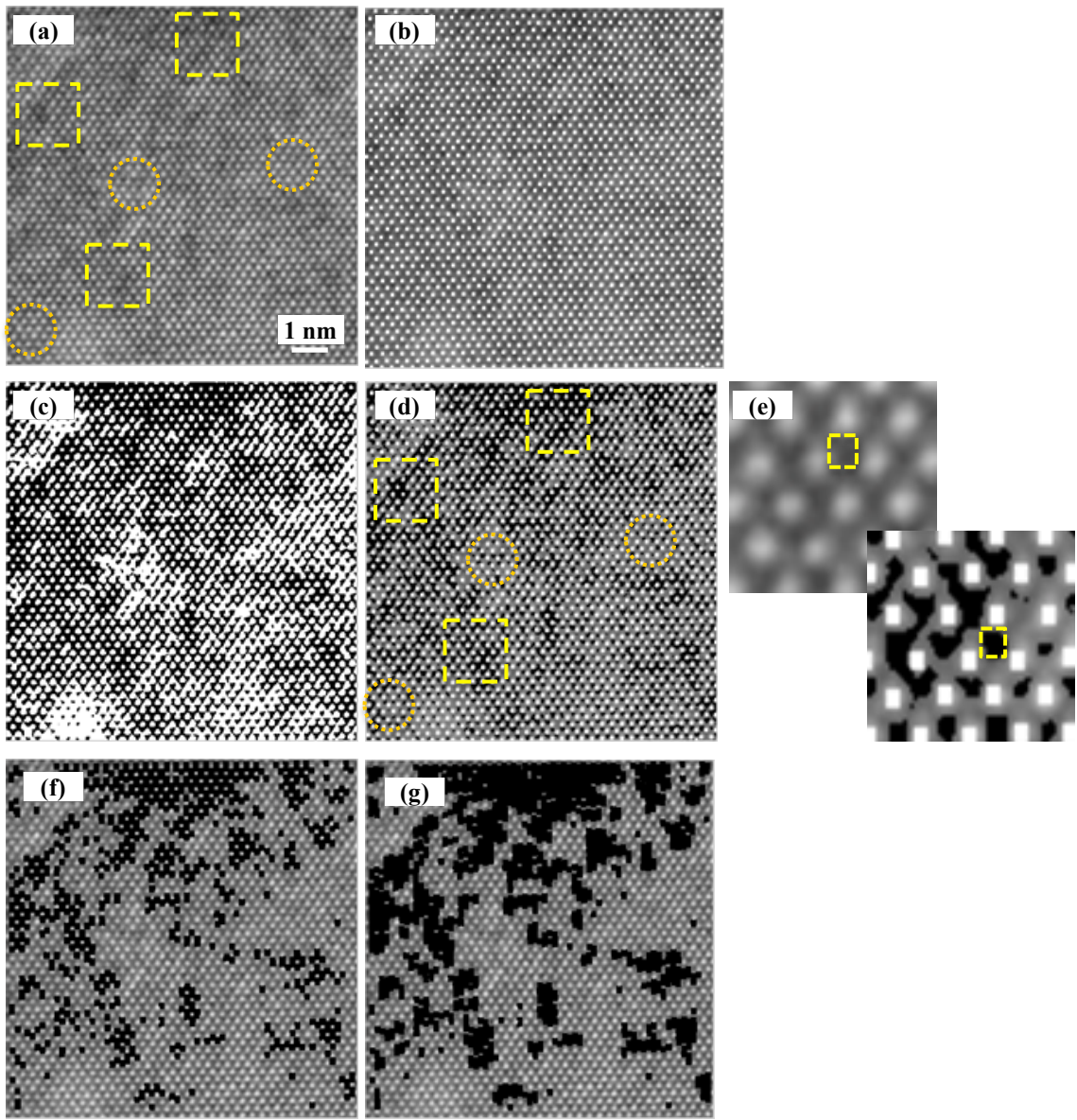

**Figure 3:** (a) An original LAADF STEM image containing BSDs; (b)-(f) Steps in the image processing to identify BSDs, as described in the main text; (g) The final image with identified BSDs marked black.

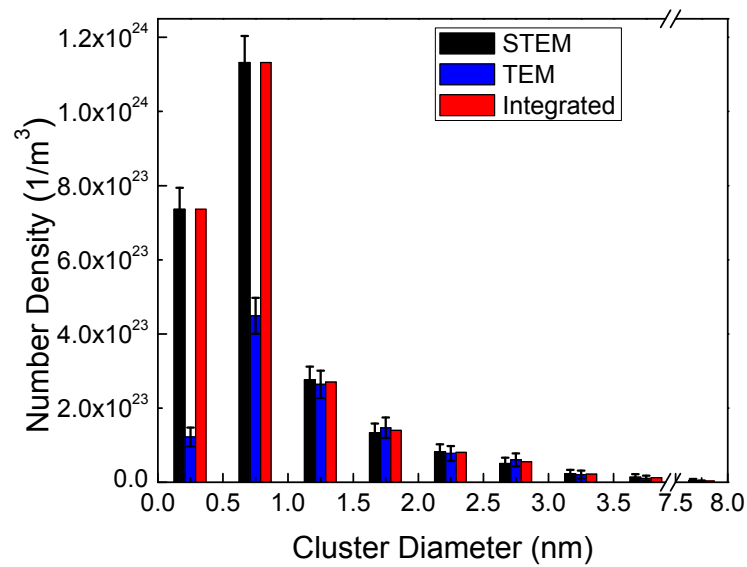

**Figure 4:** (Color online) Defect number density as a function of diameter determined by bright-field TEM, low-angle annular dark-field STEM, and by a combination of the two techniques (the integrated distribution, see Section 3.1).

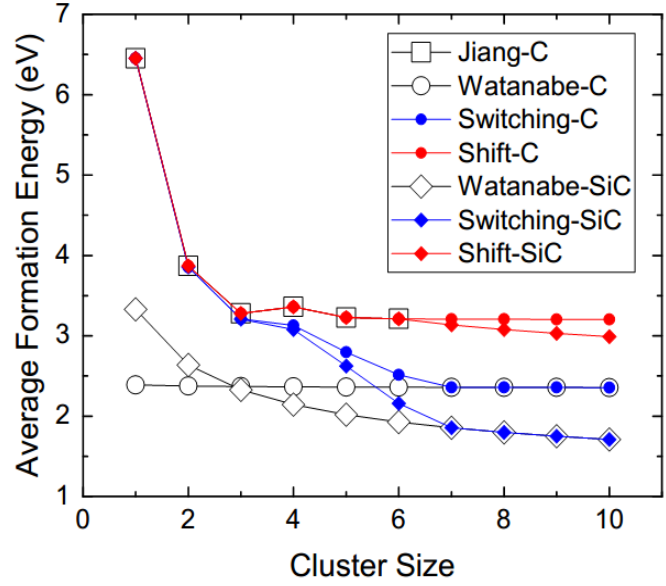

**Figure 5**: (Color online) Cluster formation energies of carbon and stoichiometric clusters. The reference state in all these calculations is a perfect 3C-SiC. DFT data (open squares), $E_f^{DFT}(n)$, is taken from Jiang *et al.* [31], empirical potential data (open circles and open diamonds), $E_f^{MD}(n)$, is taken from Watanabe *et al.* [34]. Red lines represent combined $E_f^{DFT}(n)$ and $E_f^{MD}(n)$ using the shift method, and blue lines, $E_f^{switching}(n)$ are combined $E_f^{DFT}(n)$ and $E_f^{MD}(n)$ using the switching method.

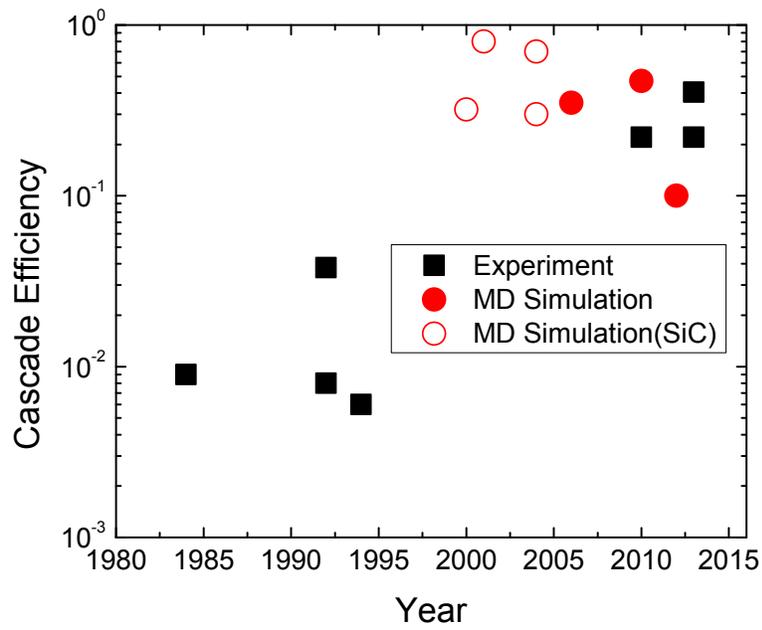

**Figure 6:** (Color online) Literature values for cascade efficiencies η in various materials determined in experiments (black squares) [37-43] and MD simulations (red circles) [44-49] Open circles are data from MD simulations on SiC [44-46].

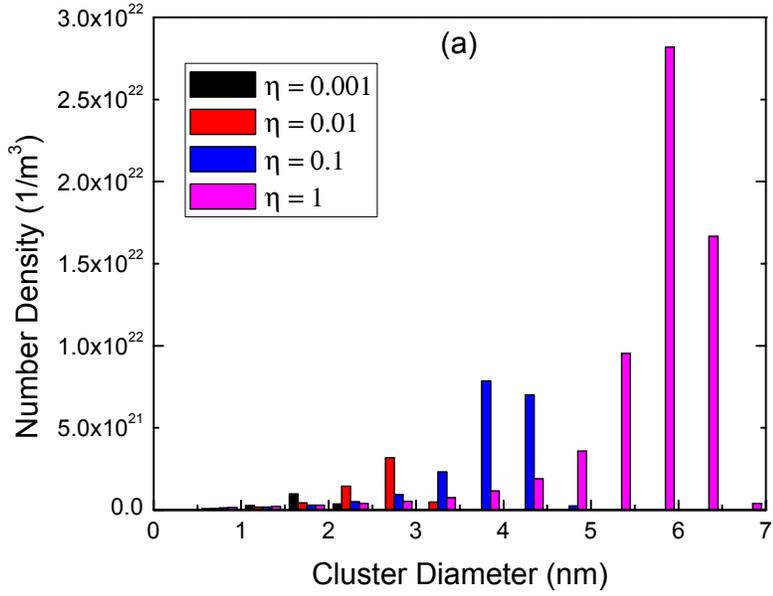

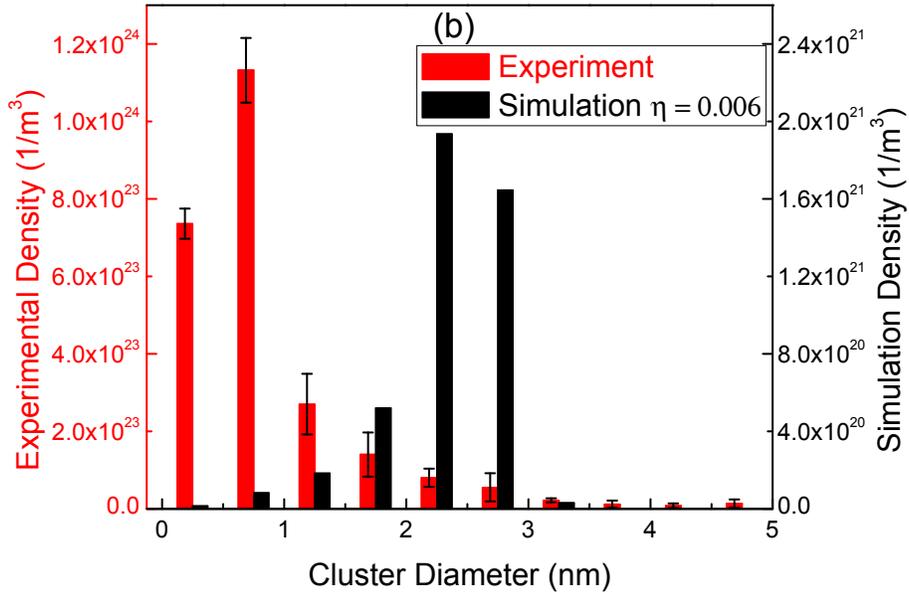

**Figure 7:** (Color online) (a) Simulation results from a CD model that includes known physics of radiation-induced defects in SiC and different cascade efficiencies, $\eta$. (b) Simulation results with the smallest reported cascade efficiency and cluster densities measured in STEM and TEM experiments.

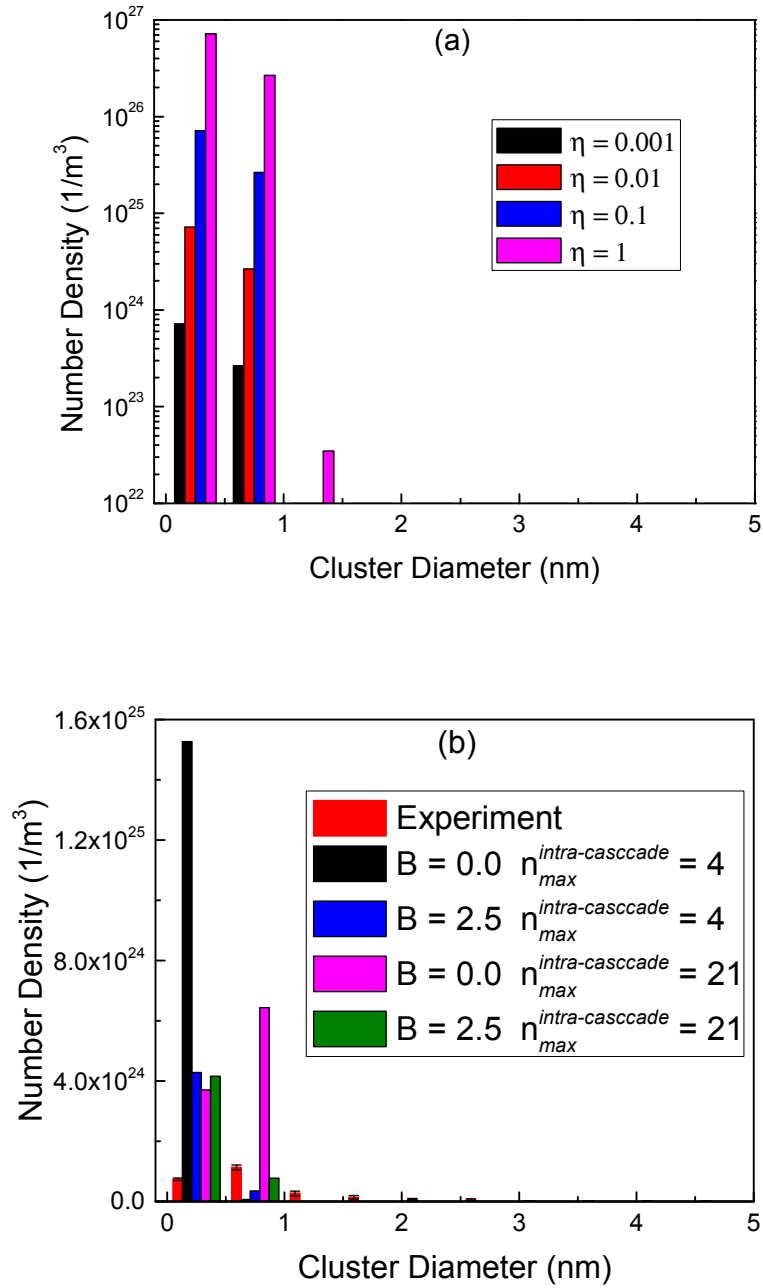

**Figure 8:** (Color online) (a) Simulation results from a CD model that includes intra-cascade cluster production. (b) Comparison of experimental number densities and those predicted by CD models that include different intra-cascade cluster production functions, $\xi_n \propto n^{-B}$, where B varies from 0.0 to 2.5 and the largest intra-cascade cluster size $n_{max}^{intra-cascade}$ varies from 4 to 21.

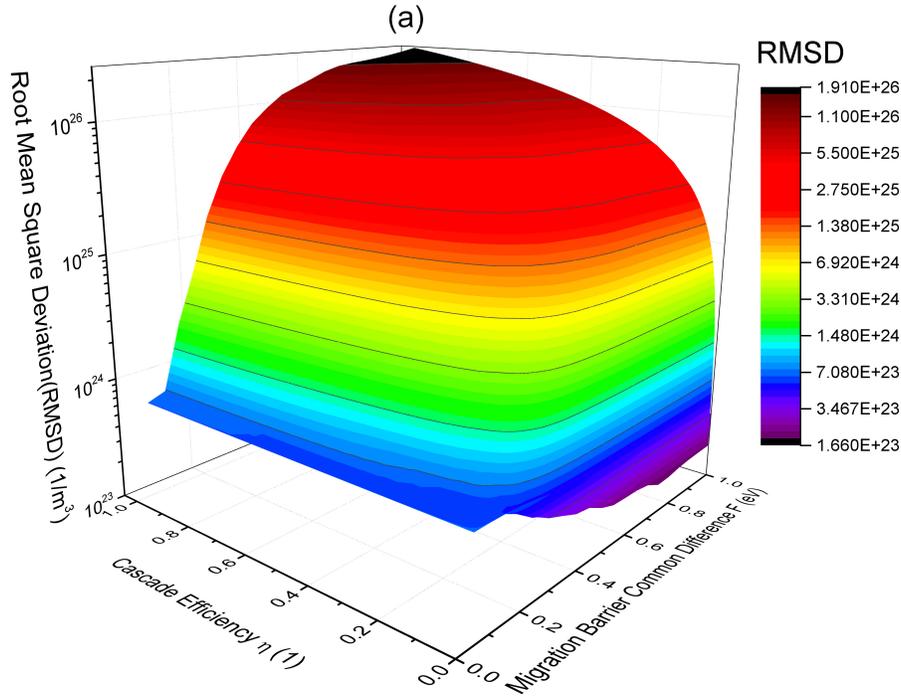

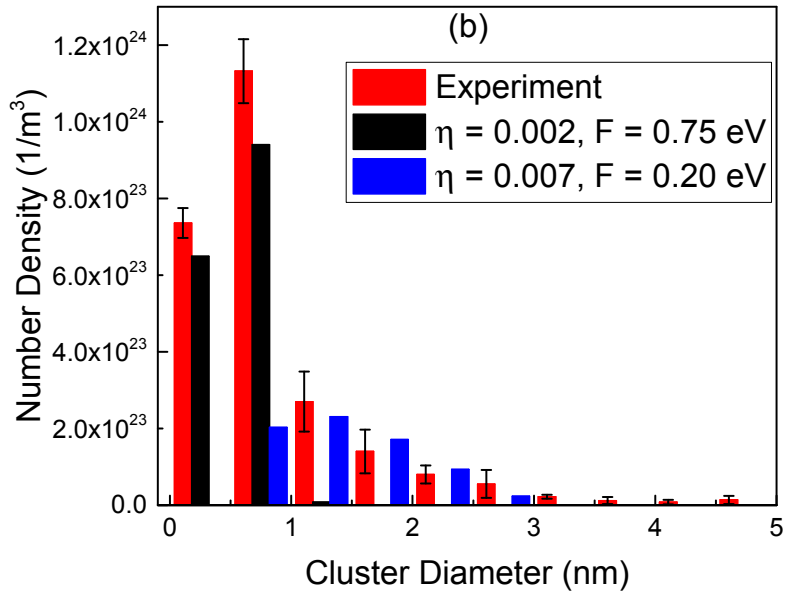

**Figure 9:** (Color online) (a) Root mean square deviation (RMSD) between experimental and simulation results based on a CD model that includes intra-cascade cluster production and cluster mobility. RMSD is plotted as a function of cascade efficiency $\eta$ and as a function of parameter $F$ in the cluster migration barrier function (see Equation 8). (b) Cluster size distributions from experiment and CD simulations. Values of $\eta = 0.002$ and $F = 0.75$ eV in the model give the smallest global RMSD. Values of $\eta = 0.007$ and $F = 0.20$ eV give the global smallest RMSD when RMSD is only calculated for clusters with diameters between 1.0 nm and 2.5 nm.

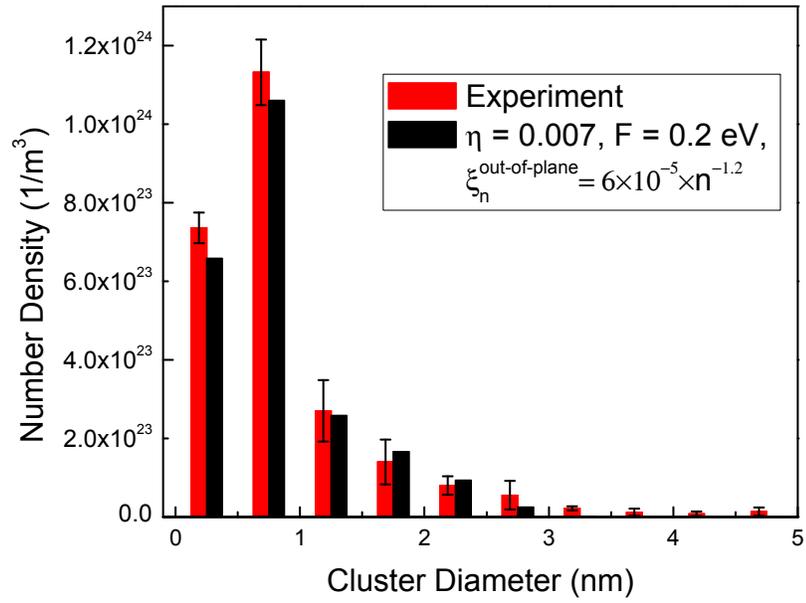

**Figure 10:** (Color online) Comparison of results from TEM and STEM experiments and from simulation based on CD model that includes intra-cascade cluster production, mobile clusters, and out-of-plane cluster production.

**Table 1:** Formation energies of PDs. DFT stands for density functional theory, MD and MS for molecular dynamics and molecular statics, respectively.

| Reference: | Method | Formation Energies (eV) |
|---|---|---|
| Shrader et al. [19] | DFT | Si rich: $E_f^{C_I} = 6.95$;   $E_f^{Si_I} = 8.75$;   $E_f^{Si_C} = 3.56$;   $E_f^{C_{Si}} = 4.03$;   C rich: $E_f^{C_I} = 6.51$;   $E_f^{Si_I} = 9.20$; |
| Jiang et al. [31] | DFT | C rich: $E_f^{C_I} = 6.45$; |
| Bockstedte et al. [18] | DFT | Si rich:   $E_f^{C_I} = 6.71$;   $E_f^{Si_I} = 8.56$; |
| Watanabe et al. [33] | MD & MS | $E_f^{C_I} = 3.24$;   $E_f^{Si_I} = 3.17$;* |
| Li et al. [32] | MD | $E_f^{C_I} = 13.67$;   $E_f^{Si_I} = 14.98$;** |

* Cohesive energies of carbon and silicon in 3C-SiC are – 6.61 eV and – 6.21 eV

** Cohesive energies of carbon and silicon in 3C-SiC are the same – 6.39 eV

**Table 2:** PDs reactions, reaction barriers and reaction capture radius implemented in our CD model. These values and details about how they are calculated and chosen can be found from Ref.[8, 35].

| Reaction # | Defect Reaction | Barrier(eV) | Radius(Å) |
|---|---|---|---|
| R.1 | $V_C + C_i \rightarrow C_C$ | 0.9 | 4.31 |
| R.2 | $V_{Si} + Si_i \rightarrow Si_{Si}$ | 0.829 | 3.08 |
| R.3 | $V_C + Si_i \rightarrow Si_C$ | 1.11 | 3.7 |
| R.4 | $V_{Si} + C_i \rightarrow C_{Si}$ | 1.25 | 3.3 |
| R.5 | $Si_i + C_{Si} \rightarrow Si_{Si} + C_i$ | 0.829 | 3.08 |
| R.6 | $C_i + Si_C \rightarrow C_C + Si_i$ | 1.34 | 4.31 |
| R.7 | $V_{Si} \leftrightarrow V_C - C_{Si}$ | 2.70/2.40 | -/- |
| R.8 | $V_C - C_{Si} \leftrightarrow V_C + C_{Si}$ | 7.33/1.25 | -/3.3 |
| R.9 | $C_i + V_C - C_{Si} \rightarrow C_C + C_{Si}$ | 0.80 | 4.7 |
| R.10 | $Si_i + V_C - C_{Si} \rightarrow Si_C + C_{Si}$ | 3.12 | 1.6 |

**Table 3:** Ratio between the number of isolated PDs and the number of interstitials trapped in clusters, as well as formation volumes of defects. Formation volumes are taking from Ref [32] and the method to calculate the ratios is explained in text.

|  | $C_I$ | $Si_I$ | $V_C$ | $V_{Si}$ | $C_{Si}$ | $Si_C$ |
|---|---|---|---|---|---|---|
| Ratio of isolated PD to cluster-trapped interstitials | $0.258 \times 10^{-8}$ | $0.110 \times 10^{-9}$ | 0.599 | 0.246 | 0.240 | 0.414 |
| Formation Volume ($\text{Å}^3$) | 15.33 | 37.22 | 2.68 | 1.85 | -9.52 | 15.44 |